\documentclass{article}
\usepackage[a4paper, left=1.5cm, right=1.5cm, top=2.5cm, bottom=2.5cm]{geometry}
\usepackage{graphicx} 
\usepackage{braket}
\usepackage{amsmath}
\usepackage{amssymb}
\usepackage{xcolor}
\usepackage{authblk} 
\usepackage{cite}

\title{Does Gravity Render Probability Quasilocal?}

\author[]{Oem Trivedi\thanks{Email: oem.trivedi@vanderbilt.edu}}

\affil[]{Department of Physics and Astronomy, Vanderbilt University, Nashville, TN 37235, USA}

\date{\today}

\begin{document}

\maketitle

\begin{abstract}
We propose that probability in quantum theory, like energy in general relativity, acquires a fundamentally quasilocal character in curved spacetime. Interpreting Hermiticity as the symmetry associated with inner-product conservation, we show that gravitational boundaries and horizons convert global probability conservation into a flux balance law. The resulting quasilocal probability naturally induces effective non-Hermiticity for restricted observers while preserving global unitarity. We demonstrate this explicitly in Schwarzschild, Kerr and FLRW spacetimes and after this, we identify observational imprints in black hole ringdowns. Our results suggest that in quantum field theory on curved backgrounds, probability conservation is as geometrically conditioned as energy itself.
\end{abstract}

\section{Introduction}

Hermiticity has traditionally been regarded as one of the core structural pillars of quantum mechanics as in the conventional Hilbert space formulation, imposing that physical observables correspond to Hermitian operators is what ensures real measurement outcomes via spectral reality, guarantees unitary time evolution for closed systems and is a core of the probabilistic interpretation through the Born rule and the conservation of the canonical inner product \cite{qm1bohm2013quantum,qm2zettili2009quantum,qm3sakurai2020modern,qm4griffiths2018introduction,qm5shankar2012principles,qm6scherrer2024quantum}. In this standard picture, Hermiticity is therefore not merely a mathematical convenience but is interwoven with the operational content of measurement theory, probability conservation and the consistency of quantum dynamics. Yet over the past several decades it has become increasingly clear that many physically relevant quantum systems are effectively open, which means that degrees of freedom may be deliberately eliminated by coarse-graining, weakly monitored or rendered permanently inaccessible by environmental coupling, dissipation or some form of kinematic constraints \cite{nh1moiseyev2011non,nh2ashida2020non,nh3hatano1996localization}. In such settings, non-unitary generators, complex potentials and gain-loss structures arise naturally as effective descriptions. This has led to non-Hermitian quantum mechanics has maturing into a systematic framework that is both theoretically rich and experimentally testable. These ideas have been realized in engineered platforms where controlled amplification, absorption and post selected dynamics permit precision exploration of regimes beyond strictly unitary evolution. Accordingly, non-Hermitian models are now studied both as effective theories of open-system dynamics and in some formulations, as generalized quantum frameworks in which probabilistic consistency is maintained through modified inner products, metric operators and biorthogonal constructions \cite{nh4jones2014relativistic,nh5gopalakrishnan2021entanglement,nh6hatano1997vortex,nh7bender2007making,nh8longhi2010optical}. This broad experimental and conceptual expansion has also motivated increasingly sensitive efforts to constrain effective non-Hermitian generators through interferometric tests, spectral-response and pseudospectral analyses, decay-profile measurements and direct consistency checks of norm conservation in appropriately defined inner products \cite{nh9jones2010non,nh10krejvcivrik2015pseudospectra,nh11cui2012geometric,nh12bergholtz2021exceptional}.
\\
\\
Recent work \cite{hermTrivedi:2026qof} has shown that Hermiticity may be a symmetry associated with the conservation of the quantum inner product, rather than a purely axiomatic requirement. The idea here is that when the relevant inner-product current is globally conserved and no boundary flux is present, the canonical Hermiticity condition emerges as the corresponding symmetry constraint on the generator of evolution. In this work we take this one step further and establish how gravity and the causal structure of spacetime can fundamentally alter the implementation of probability conservation, motivating a quasilocal description of normalization and inner-product charge for restricted observers and regions. In the next section we give a slight overview of non-Hermiticity in quantum mechanics, while in Section 3 we review the recently developed idea of Hermiticity as a symmetry law. We then propose the idea of a "Quasilocal Probability" in Section 4, and show how it explicitly could manifest in Schwarzschild and FLRW spacetimes as examples in Section 5. We discuss if there could be observational imprints of all this in section 6 and then conclude our work in Section 7.

\section{Non-Hermiticity in Quantum Mechanics}
One can extend standard quantum mechanics by relaxing the requirement that the Hamiltonian be self-adjoint with respect to the canonical Hilbert space inner product while preserving Schr\"odinger time evolution,
\begin{equation}
i\hbar \frac{\partial}{\partial t}\ket{\psi(t)}=\hat H \ket{\psi(t)}
\label{eq:NH_Schro}
\end{equation}
In the conventional formulation Hermiticity of $\hat H$ guarantees real spectra and unitary evolution whereas allowing $\hat H\neq \hat H^{\dagger}$ modifies the dynamical structure at its foundation. Any linear operator admits the decomposition
\begin{equation}
\hat H=\hat H_{\rm H}+i\hat \Gamma
\label{eq:H_decomp}
\end{equation}
with $\hat H_{\rm H}=\hat H_{\rm H}^{\dagger}$ and $\hat \Gamma=\hat \Gamma^{\dagger}$. The Hermitian part governs coherent oscillatory evolution while the anti-Hermitian component parametrizes dissipative or amplifying behavior. Using Eq.\eqref{eq:NH_Schro} and its adjoint the time derivative of the canonical norm becomes
\begin{equation}
\frac{d}{dt}\braket{\psi(t)|\psi(t)}
=
\frac{i}{\hbar}\bra{\psi(t)}\big(\hat H^{\dagger}-\hat H\big)\ket{\psi(t)}
=
-\frac{2}{\hbar}\bra{\psi(t)}\hat \Gamma\ket{\psi(t)}
\label{eq:norm_evol}
\end{equation}
Hence unless $\hat \Gamma=0$ the norm is not conserved and the expectation value of $\hat \Gamma$ determines exponential decay or amplification rates. Such structures arise naturally in effective descriptions of unstable states optical potentials resonance phenomena and open quantum systems \cite{nh1moiseyev2011non,nh2ashida2020non,nh3hatano1996localization}.
\\
\\
The spectral problem for a non-Hermitian Hamiltonian differs qualitatively from the Hermitian case. Right eigenstates satisfy
\begin{equation}
\hat H\ket{n_{\rm R}}=E_n\ket{n_{\rm R}}
\end{equation}
with generically complex eigenvalues $E_n=E_n^{\rm R}+iE_n^{\rm I}$. Time evolution of such a mode yields
\begin{equation}
e^{-iE_n t/\hbar}
=
e^{-iE_n^{\rm R}t/\hbar}
e^{E_n^{\rm I}t/\hbar}
\end{equation}
so the imaginary part of the spectrum controls growth or decay. Because $\hat H$ and $\hat H^{\dagger}$ do not share eigenvectors one introduces a biorthogonal structure
\begin{equation}
\hat H\ket{n_{\rm R}}=E_n\ket{n_{\rm R}}
\qquad
\hat H^{\dagger}\ket{n_{\rm L}}=E_n^{\ast}\ket{n_{\rm L}}
\qquad
\braket{n_{\rm L}|m_{\rm R}}=\delta_{nm}
\label{eq:biorth}
\end{equation}
with completeness $\sum_n \ket{n_{\rm R}}\bra{n_{\rm L}}=\mathbb{I}$. Expectation values are then evaluated as $\bra{\psi_{\rm L}}\hat O\ket{\psi_{\rm R}}$ ensuring internal spectral consistency even for complex eigenvalues \cite{nh4jones2014relativistic,nh5gopalakrishnan2021entanglement,nh6hatano1997vortex,nh7bender2007making,nh8longhi2010optical}.
\\
\\
A distinguished subclass of non-Hermitian theories admits a generalized notion of unitarity. If there exists a positive definite metric operator $\eta$ such that
\begin{equation}
\hat H^{\dagger}=\eta \hat H \eta^{-1}
\end{equation}
then $\hat H$ is pseudo-Hermitian and the modified inner product
\begin{equation}
\langle\psi|\phi\rangle_{\eta}=\langle\psi|\eta|\phi\rangle
\end{equation}
is conserved provided $\eta$ is time independent,
\begin{equation}
\frac{d}{dt}\langle\psi(t)|\psi(t)\rangle_{\eta}=0
\end{equation}
This structure underlies $\mathcal{PT}$-symmetric quantum mechanics and related frameworks in which real spectra and probabilistic consistency are retained in a generalized Hilbert space geometry \cite{nh7bender2007making,nh8longhi2010optical}. More broadly non-Hermitian Hamiltonians appear as effective generators in open systems non-equilibrium settings and engineered gain-loss platforms where exceptional points and sensitivity enhancement have been experimentally explored \cite{nh13gardas2016non,nh14matsoukas2023non,nh15bender2007faster,nh16cao2023statistical,nh17giri2009non,nh18ju2019non,nh19ju2024emergent}. In a recent work \cite{Trivedi:2026itv} it was shown that effective non-Hermiticity at a global level can itself be constrained by cosmological considerations within quantum cosmology frameworks \cite{qc1bojowald2015quantum,qc2bojowald2008loop,qc3misner1969quantum,qc4wiltshire1996introduction,qc5ashtekar2011loop,qc6hawking1987quantum,qc7gell1996quantum,qc8bojowald2011quantum,qc9vilenkin1995predictions,qc10calcagni2017classical}. In this sense the anti-Hermitian component $\hat \Gamma$ provides a quantitative measure of probability flux irreversibility and coarse-grained dynamics bridging microscopic generators with experimentally measurable decay rates amplification factors and spectral shifts.

\section{Hermiticity as a Symmetry and Gravity}
In the conventional formulation of quantum mechanics Hermiticity is imposed as a structural requirement on the Hamiltonian so that
\begin{equation}
\hat H=\hat H^{\dagger}
\end{equation}
with respect to the canonical Hilbert space inner product. In a recent work \cite{hermTrivedi:2026qof}, this condition was reinterpreted as the manifestation of a deeper symmetry principle associated with conservation of the quantum inner product itself. Consider a generalized inner product defined by a positive operator $\eta$,
\begin{equation}
\langle\psi|\phi\rangle_{\eta}=\langle\psi|\eta|\phi\rangle
\end{equation}
and assume Schr\"odinger evolution
\begin{equation}
i\hbar \frac{\partial}{\partial t}\ket{\psi(t)}=\hat H \ket{\psi(t)}
\end{equation}
Then demanding conservation of the $\eta$-norm,
\begin{equation}
\frac{d}{dt}\langle\psi(t)|\psi(t)\rangle_{\eta}=0
\end{equation}
leads to the operator identity
\begin{equation}
\hat H^{\dagger}\eta-\eta \hat H=i\hbar \dot\eta
\end{equation}
If $\eta$ is time independent this reduces to
\begin{equation}
\hat H^{\dagger}\eta=\eta \hat H
\end{equation}
which expresses Hermiticity as a symmetry condition of the inner product structure. In the canonical case $\eta=\mathbb I$ one recovers $\hat H=\hat H^{\dagger}$ and thus Hermiticity emerges not as an isolated postulate but as the statement that the inner-product charge associated with the chosen metric on Hilbert space is conserved under time evolution.
\\
\\
The same structure admits a covariant formulation in quantum field theory as for a complex scalar field the Klein Gordon equation implies conservation of the bilinear current
\begin{equation}
J^{\mu}(\Phi_1,\Phi_2)
=
-i\left(\Phi_1^{\ast}\nabla^{\mu}\Phi_2-\Phi_2\nabla^{\mu}\Phi_1^{\ast}\right)
\end{equation}
with
\begin{equation}
\nabla_{\mu}J^{\mu}=0
\end{equation}
On a spacelike hypersurface $\Sigma$ with unit normal $n^{\mu}$ the associated inner product is
\begin{equation}
(\Phi_1,\Phi_2)_{\Sigma}
=
\int_{\Sigma} d\Sigma \, n_{\mu} J^{\mu}(\Phi_1,\Phi_2)
\end{equation}
and conservation of $J^{\mu}$ ensures independence of the choice of Cauchy surface provided no boundary flux is present. In this language Hermiticity of the generator of evolution is equivalent to the statement that the symplectic or inner-product form on the space of solutions is preserved by time translations. The existence of a globally well defined conserved charge therefore depends not only on the local field equations but also on global properties of spacetime such as hyperbolicity and the absence of flux through boundaries.
\\
\\
Gravity enters this structure in a fundamental way. In a curved spacetime with metric $g_{\mu\nu}$ the conservation equation $\nabla_{\mu}J^{\mu}=0$ remains valid on shell but the associated charge depends on the geometry of the hypersurface and on the causal structure of the manifold. In particular, if one considers a spacetime region with boundary the divergence theorem yields
\begin{equation}
Q_{R}[\Sigma_2]-Q_{R}[\Sigma_1]
=
-\int_{B_{R}} d\Sigma_{\mu} J^{\mu}
\end{equation}
where $Q_{R}$ is the inner-product charge restricted to a spatial region $R$ and $B_{R}$ denotes the worldtube generated by its boundary. Thus conservation of the inner product becomes a flux balance law once gravitationally induced horizons or finite causal domains are present. This perspective establishes a structural link between inner-product conservation and gravitational geometry as the Einstein equations imply
\begin{equation}
\nabla_{\mu}T^{\mu\nu}=0
\end{equation}
through the contracted Bianchi identity
\begin{equation}
\nabla_{\mu}G^{\mu\nu}=0
\end{equation}
ensuring consistency of energy flux in curved spacetime. In stationary geometries admitting a timelike Killing vector $\xi^{\mu}$ one may construct a conserved energy current
\begin{equation} \label{kill}
j^{\mu}_{(\xi)}=T^{\mu\nu}\xi_{\nu}
\end{equation}
with $\nabla_{\mu}j^{\mu}_{(\xi)}=0$, leading to quasilocal notions of mass and energy when boundaries are present. The inner-product current obeys an analogous divergence equation and in the presence of horizons or restricted observer algebras its associated charge could likewise become quasilocal. In this way Hermiticity can be understood as the global realization of a symmetry associated with inner-product conservation, while gravity through its control of causal domains and boundary structure determines whether such conservation can be implemented globally or only quasilocally.

\section{Quasilocal Probabilities}
The discussion of the previous sections suggests that probability conservation in quantum theory is most naturally understood as the conservation of an inner-product charge associated with a divergence-free current. In a relativistic field theoretic setting this statement is encoded in
\begin{equation}
\nabla_{\mu}J^{\mu}=0
\end{equation}
where $J^{\mu}$ denotes the bilinear inner-product current. On a spacelike hypersurface $\Sigma$ with unit normal $n^{\mu}$ the associated charge is
\begin{equation}
Q[\Sigma]=\int_{\Sigma} d\Sigma \, n_{\mu} J^{\mu}
\end{equation}
and in the absence of boundaries or flux at infinity the divergence theorem guarantees that $Q[\Sigma]$ is independent of the choice of Cauchy slice. In this global setting the generator of time evolution is Hermitian with respect to the induced inner product and probability is conserved in the usual sense. The situation changes when one restricts attention to a spatial region $R\subset\Sigma$ as then the natural generalization of the inner-product charge is then
\begin{equation}
Q_{R}[\Sigma]=\int_{\Sigma\cap R} d\Sigma \, n_{\mu} J^{\mu}
\end{equation}
Applying the divergence theorem to the spacetime slab bounded by $\Sigma_{1}\cap R$, $\Sigma_{2}\cap R$ and the timelike worldtube $B_{R}$ generated by $\partial R$ yields the quasilocal balance law
\begin{equation}
Q_{R}[\Sigma_{2}]-Q_{R}[\Sigma_{1}]
=
-\int_{B_{R}} d\Sigma_{\mu} J^{\mu}
\label{eq:quasilocal_balance}
\end{equation}
This equation provides a precise definition of quasilocal probability. The quantity $Q_{R}$ measures the normalization or inner-product charge contained within a finite region and its time variation is governed entirely by flux through the boundary. Probability conservation is therefore no longer an absolute global statement but a balance condition relating bulk and boundary contributions.
\\
\\
The connection to non-Hermiticity follows directly because if evolution in the region $R$ is written as
\begin{equation} \label{evol}
i\hbar \frac{\partial}{\partial t}\Phi = H_{R}\Phi
\end{equation}
with inner product defined by $Q_{R}$, then
\begin{equation} 
\frac{d}{dt}(\Phi,\Phi)_{R}
=
i(\Phi,(H_{R}-H_{R}^{\dagger})\Phi)_{R}
\end{equation}
Comparing with Eq. \eqref{eq:quasilocal_balance} gives
\begin{equation} \label{bala}
i(\Phi,(H_{R}-H_{R}^{\dagger})\Phi)_{R}
=
-\int_{\partial R} dS \, J^{\mu}s_{\mu}
\end{equation}
where $s_{\mu}$ is the outward pointing normal to the boundary. The anti-Hermitian part of the regional generator is therefore proportional to the flux of inner-product current through $\partial R$. Quasilocal probability and effective non-Hermiticity are two descriptions of the same boundary phenomenon. Whenever the background spacetime or the observer’s causal domain introduces a nontrivial boundary, probability conservation within that domain becomes conditional upon vanishing flux. In flat Minkowski space with $R=\mathbb{R}^{3}$ and suitable falloff conditions the boundary integral vanishes and global Hermiticity is recovered. In curved spacetimes containing horizons or finite causal patches the boundary is intrinsic and quasilocality becomes physically unavoidable. This structure closely parallels the quasilocal description of energy in general relativity as we know that given a Killing vector field $\xi^{\mu}$,
one can construct a conserved energy current given by \eqref{kill} and the associated energy contained in a region $R$ satisfies a flux balance law analogous to Eq.\eqref{eq:quasilocal_balance}. In the presence of boundaries or horizons one obtains quasilocal notions of energy such as ADM or Misner-Sharp mass, whose time variation is determined by stress-energy flux through the boundary.
\\
\\
There is therefore a structural correspondence as hermiticity symmetry in quantum theory ensures conservation of the inner-product charge in precisely the same way that Killing symmetry ensures conservation of energy. In the first case the relevant condition may be written as
\begin{equation}
H^{\dagger}\eta = \eta H
\end{equation}
or in the canonical case $H=H^{\dagger}$, expressing invariance of the inner product under time evolution. In the second case the Killing equation $\nabla_{(\mu}\xi_{\nu)}=0$ expresses invariance of the metric under the flow generated by $\xi^{\mu}$. Both conditions encode a symmetry that leads to a conserved current. When such symmetry is globally implementable the corresponding charge is globally conserved. When the symmetry is obstructed by geometry or by restriction to a subregion the associated charge becomes quasilocal and obeys a flux balance law.
\\
\\
The analogy is not exact as energy is sourced by stress tensor components and its quasilocal definitions depend explicitly on gravitational field equations. Quasilocal probability arises from the inner-product current and depends only on the quantum dynamical structure. Nevertheless both share a common geometric form as integrals over hypersurfaces whose variation is governed by boundary flux. This similarity suggests that the quasilocalization of probability in curved spacetime is as natural as the quasilocalization of energy in general relativity.
\begin{table}[t]
\centering
\begin{tabular}{|c|c|}
\hline
Quasilocal Probability & Quasilocal Energy \\
\hline
Current $J^{\mu}$ with $\nabla_{\mu}J^{\mu}=0$ & Current $j^{\mu}_{(\xi)}$ with $\nabla_{\mu}j^{\mu}_{(\xi)}=0$ \\
\hline
Charge $Q_{R}=\int_{\Sigma\cap R} n_{\mu}J^{\mu}$ & Energy $E_{R}=\int_{\Sigma\cap R} n_{\mu}j^{\mu}_{(\xi)}$ \\
\hline
Flux through $\partial R$ controls $H_{R}-H_{R}^{\dagger}$ & Flux through $\partial R$ controls change of mass \\
\hline
Hermiticity symmetry ensures global conservation & Killing symmetry ensures global conservation \\
\hline
Obstruction by horizons yields effective non-Hermiticity & Obstruction by horizons yields quasilocal mass concepts \\
\hline
\end{tabular}
\caption{Conceptual correspondence between quasilocal probability in quantum field theory and quasilocal energy in general relativity. In both cases a conserved current defines a charge within a finite region $R$, and boundary flux through $\partial R$ governs the deviation from global conservation.}
\label{tab:quasilocal_comparison}
\end{table}
In this formulation probability conservation becomes a geometric statement sensitive to the causal and boundary structure of spacetime. Quasilocal probability thus provides a natural extension of the Hermiticity symmetry principle into curved backgrounds and sets the stage for explicit realizations in Schwarzschild and FLRW geometries. THe comparison of quasilocal energy and quasilocal probability is summarized in table \ref{tab:quasilocal_comparison}.
\\
\\
Also, quasilocal probability is deeply intertwined with the notion of information because it quantifies how much of a quantum state’s norm, therefore how much of its informational content, is accessible within a given spacetime region. When the inner-product current satisfies $\nabla_\mu J^\mu=0$, global probability conservation is guaranteed but restricting to a finite domain converts this into a flux balance law. The boundary flux then measures the flow of quantum information across causal horizons or spatial boundaries. In this sense, quasilocal probability is not merely a normalization issue but a statement about how information is partitioned between accessible and inaccessible degrees of freedom. Effective non-Hermiticity in a restricted region reflects the loss or gain of information due to this geometric flux, closely paralleling how entanglement entropy grows when correlations are traced out. Thus quasilocal probability provides a geometric bridge between probability conservation, information flow and the causal structure of spacetime, suggesting that in curved backgrounds the distribution of quantum information is as fundamentally constrained by geometry as energy is.

\section{Quasilocal Probabilities in Schwarzschild and FLRW Spacetimes}
We now explicitly construct the quasilocal inner product charge in two representative curved spacetimes and demonstrate how the metric structure directly feeds into effective non Hermiticity for restricted regions. Throughout we consider a complex Klein Gordon field satisfying
\begin{equation} \label{kg}
\Box_g \Phi - m^2 \Phi = 0
\end{equation}
with conserved current
\begin{equation} \label{cc}
J^{\mu} = - i \left( \Phi^{\ast} \nabla^{\mu} \Phi - \Phi \nabla^{\mu} \Phi^{\ast} \right)
\end{equation}
and
\begin{equation}
\nabla_{\mu} J^{\mu} = 0
\end{equation}
The quasilocal inner product on a spacelike hypersurface $\Sigma$ restricted to a region $R$ is
\begin{equation}
Q_R[\Sigma] = \int_{\Sigma \cap R} d\Sigma \, n_{\mu} J^{\mu}
\end{equation}
where $n^{\mu}$ is the unit normal to $\Sigma$ and $d\Sigma = \sqrt{h} \, d^3 x$ with $h$ the determinant of the induced spatial metric.
\\
\\
Consider the Schwarzschild geometry
\begin{equation}
ds^2 = - f(r) dt^2 + \frac{dr^2}{f(r)} + r^2 d\Omega^2
\end{equation}
with
\begin{equation}
f(r) = 1 - \frac{2GM}{r}
\end{equation}
On a constant time hypersurface $\Sigma_t$ the unit normal is
\begin{equation}
n^{\mu} = \left( \frac{1}{\sqrt{f}} , 0 , 0 , 0 \right)
\end{equation}
The induced metric determinant is
\begin{equation}
\sqrt{h} = \frac{r^2 \sin\theta}{\sqrt{f}}
\end{equation}
Hence the quasilocal charge in a radial shell $r \in [ r_1 , r_2 ]$ becomes
\begin{equation}
Q_{[ r_1 , r_2 ]} (t) = \int_{r_1}^{r_2} dr \int d\Omega \, r^2 J^t
\end{equation}
All dependence on $f(r)$ cancels in the measure. Applying the divergence theorem to the spacetime slab bounded by $t_1$ and $t_2$ gives
\begin{equation}
\frac{d}{dt} Q_{[ r_1 , r_2 ]} (t) = - \int d\Omega \, r^2 J^r \Big|_{r_1}^{r_2}
\end{equation}
Now choose $R$ to be the exterior region $r \ge r_h + \epsilon$ with $r_h = 2GM$ and assume vanishing flux at spatial infinity. Then
\begin{equation}
\frac{d}{dt} Q_{\text{ext}} (t) = - \int d\Omega \, r^2 J^r \Big|_{r = r_h + \epsilon}
\end{equation}
If evolution in the exterior is written as $i \partial_t \Phi = H_{\text{ext}} \Phi$ then
\begin{equation}
i ( \Phi , ( H_{\text{ext}} - H_{\text{ext}}^{\dagger} ) \Phi )_{\text{ext}} = - \int d\Omega \, r^2 J^r \Big|_{r = r_h + \epsilon}
\end{equation}
The anti Hermitian part of the exterior Hamiltonian is therefore proportional to the horizon flux.  

To make this explicit consider an ingoing mode near the horizon written in tortoise coordinate $r_\ast$
\begin{equation}
\Phi_{\text{in}} = \frac{A}{r} Y_{\ell m} e^{- i \omega ( t + r_\ast )}
\end{equation}
with $dr_\ast / dr = 1 / f$. One finds
\begin{equation}
J^{r_\ast} = - \frac{2 \omega}{f} | \Phi_{\text{in}} |^2
\end{equation}
which is negative and directed toward the horizon. Hence
\begin{equation}
\frac{d}{dt} Q_{\text{ext}} (t) \neq 0
\end{equation}
and the exterior generator is not self adjoint unless one includes the interior region. In the limit $M \to 0$ one has $f(r) \to 1$ and the horizon disappears. For the full space with standard falloff
\begin{equation}
\int_{r \to \infty} r^2 J^r d\Omega = 0
\end{equation}
so
\begin{equation}
\frac{d}{dt} Q (t) = 0
\end{equation}
and Hermiticity is restored globally. Consider now the spatially flat FLRW metric
\begin{equation}
ds^2 = - dt^2 + a(t)^2 \left( dr^2 + r^2 d\Omega^2 \right)
\end{equation}
On $t = \text{const}$ slices
\begin{equation}
n^{\mu} = ( 1 , 0 , 0 , 0 )
\end{equation}
and
\begin{equation}
\sqrt{h} = a(t)^3 r^2 \sin\theta
\end{equation}
For a comoving ball $0 \le r \le R$
\begin{equation}
Q_R (t) = \int_0^R dr \int d\Omega \, a^3 r^2 J^t
\end{equation}
The boundary worldtube at $r = R$ yields
\begin{equation}
\frac{d}{dt} Q_R (t) = - \int d\Omega \, a^3 R^2 J^r ( t , R )
\end{equation}
If $i \partial_t \Phi = H_R \Phi$ then
\begin{equation}
i ( \Phi , ( H_R - H_R^{\dagger} ) \Phi )_R = - \int d\Omega \, a^3 R^2 J^r
\end{equation}
Thus the anti Hermitian contribution is weighted by the scale factor through the spatial volume element. To illustrate consider an outgoing spherical wave in conformal time $\eta$
\begin{equation}
\Phi_{\text{out}} = \frac{A}{a(\eta) r} Y_{\ell m} e^{- i \omega ( \eta - r )}
\end{equation}
Using $g^{rr} = 1 / a^2$ one obtains
\begin{equation}
J^r = \frac{2 \omega |A|^2}{a^4 r^2} | Y_{\ell m} |^2
\end{equation}
which is positive and directed outward. Substituting into the flux expression gives
\begin{equation}
\frac{d}{dt} Q_R (t) = - \frac{2 \omega |A|^2}{a^2}
\end{equation}
after angular integration. The rate of change scales as $a^{-2}$ showing that expansion directly modulates the quasilocal probability flux. Note also that in the limit $a(t) \to 1$ the metric reduces to Minkowski space and the quasilocal charge becomes
\begin{equation}
Q_R (t) = \int_0^R dr \int d\Omega \, r^2 J^t
\end{equation}
For the entire space and appropriate falloff the boundary term at infinity vanishes and global Hermiticity is recovered.
\\
\\
These explicit constructions demonstrate that quasilocal probability is determined by the geometric measure factors $\sqrt{h}$ and by the boundary flux weighted by the metric components. In Schwarzschild spacetime the presence of a horizon introduces an intrinsic boundary whose flux produces an anti Hermitian contribution for exterior observers. In FLRW spacetime the scale factor appears explicitly in the quasilocal charge and in the boundary term so that cosmological expansion rescales the effective probability contained in finite comoving domains. The fact that the flux term carries factors of $a(t)$ implies that expansion dilutes or redshifts the regional normalization rate and therefore links probability balance directly to the dynamical geometry of spacetime. This establishes that probability conservation is globally maintained by the divergence free current but becomes quasilocal when observers are restricted by horizons or finite domains. In both spacetimes, inclusion of the complete Cauchy surface eliminates boundary flux and restores global Hermiticity, ensuring that the fundamental theory remains unitary while regional descriptions acquire effective non Hermitian structure determined entirely by geometric flux through the boundary.
\\
\\
We now extend the above construction to the Kerr geometry and then compare the Schwarzschild, Kerr and FLRW cases from the perspective of quasilocal probability. The Kerr spacetime in Boyer-Lindquist coordinates is
\begin{equation}
ds^2 = - \left( 1 - \frac{2 M r}{\Sigma} \right) dt^2 - \frac{4 M a r \sin^2\theta}{\Sigma} dt \, d\phi + \frac{\Sigma}{\Delta} dr^2 + \Sigma d\theta^2
+ \sin^2\theta \left( r^2 + a^2 + \frac{2 M a^2 r \sin^2\theta}{\Sigma} \right) d\phi^2
\end{equation}
where
\begin{equation}
\Sigma = r^2 + a^2 \cos^2\theta
\qquad
\Delta = r^2 - 2 M r + a^2
\end{equation}
and the outer horizon is located at
\begin{equation}
r_+ = M + \sqrt{M^2 - a^2}
\end{equation}
The angular velocity of the horizon is
\begin{equation}
\Omega_H = \frac{a}{r_+^2 + a^2}
\end{equation}
and the horizon generator is the Killing field
\begin{equation}
\chi^\mu = \left( \partial_t \right)^\mu + \Omega_H \left( \partial_\phi \right)^\mu
\end{equation}
As before, we consider a complex Klein-Gordon field satisfying \eqref{kg} with conserved current \eqref{cc}
\begin{equation}
J^\mu = - i \left( \Phi^\ast \nabla^\mu \Phi - \Phi \nabla^\mu \Phi^\ast \right)
\end{equation}
and
\begin{equation}
\nabla_\mu J^\mu = 0
\end{equation}
The quasilocal inner product on a hypersurface \(\Sigma\) restricted to a region \(R\) is again
\begin{equation*}
Q_R[\Sigma] = \int_{\Sigma \cap R} d\Sigma \, n_\mu J^\mu
\end{equation*}
Now, unlike Schwarzschild, Kerr is not static so the \(t=\mathrm{const}\) slices are not orthogonal to the stationary Killing flow \(\partial_t\) and it is therefore useful to write the metric in \(3+1\) form,
\begin{equation}
ds^2 = - N^2 dt^2 + h_{ij} \left( dx^i + N^i dt \right)\left( dx^j + N^j dt \right)
\end{equation}
with lapse \(N\), shift \(N^i\), and induced spatial metric \(h_{ij}\). On \(t=\mathrm{const}\) slices, the future directed unit normal is
\begin{equation}
n_\mu = (-N,0,0,0)
\qquad
n^\mu = \left( \frac{1}{N}, - \frac{N^i}{N} \right)
\end{equation}
The quasilocal charge then becomes
\begin{equation}
Q_R(t) = \int_R d^3x \, \sqrt{h}\, n_\mu J^\mu
\end{equation}
For a shell \(r \in [r_1,r_2]\), the divergence theorem gives
\begin{equation}
\frac{d}{dt} Q_{[r_1,r_2]}(t)
=
- \int_{\partial R} dS \, s_\mu J^\mu
\end{equation}
where \(s^\mu\) is the outward pointing unit normal to the timelike boundary worldtube. If we choose the exterior region \(R\) to be \(r \ge r_+ + \epsilon\) and assume vanishing flux at spatial infinity, then
\begin{equation}
\frac{d}{dt} Q_{\mathrm{ext}}(t)
=
- \int_{\mathcal H_\epsilon} dS \, s_\mu J^\mu
\end{equation}
where \(\mathcal H_\epsilon\) is the stretched horizon at \(r=r_+ + \epsilon\). Thus, exactly as in Schwarzschild, the exterior charge fails to be conserved whenever there is nonzero current into the horizon and if we write the regional evolution with \eqref{evol} and \eqref{bala} then we see that the anti-Hermitian part of the exterior generator is again determined by horizon flux.
\\
\\
The important difference from Schwarzschild appears in the structure of the near-horizon modes as for a separated mode of the form
\begin{equation*}
\Phi = e^{- i \omega t} e^{i m \phi} S_{\ell m}(\theta) R_{\ell m}(r)
\end{equation*}
Near \(r=r_+\), the ingoing solution behaves as
\begin{equation}
\Phi_{\mathrm{in}}
\sim
A \, S_{\ell m}(\theta)\, e^{- i \omega t} e^{i m \phi} e^{- i (\omega - m \Omega_H) r_\ast}
\end{equation}
It is useful to introduce the co rotating azimuthal angle
\begin{equation}
\tilde{\phi} = \phi - \Omega_H t
\end{equation}
in terms of which the mode can be rewritten as
\begin{equation}
\Phi_{\mathrm{in}}
\sim
A \, S_{\ell m}(\theta)\, e^{- i \omega t} e^{i m \tilde{\phi}} e^{- i (\omega - m \Omega_H) (t + r_\ast)}
\end{equation}
This shows immediately that the effective frequency governing horizon transport is not \(\omega\) itself but
\begin{equation}
\tilde{\omega} = \omega - m \Omega_H
\end{equation}
The reason is geometric in the sense that the natural time translation at the horizon is generated by \(\chi^\mu\), not by \(\partial_t\) alone. To see how this enters the current, consider the radial flux near the horizon where since the current is bilinear in the field and its derivative, the dominant radial contribution is controlled by the derivative of the phase with respect to \(r_\ast\) and so we have,
\begin{equation}
\partial_{r_\ast} \Phi_{\mathrm{in}}
=
- i (\omega - m \Omega_H)\, \Phi_{\mathrm{in}}
\end{equation}
and therefore the radial current takes the form
\begin{equation}
J^{r_\ast}
=
- i \left( \Phi^\ast \partial^{r_\ast} \Phi - \Phi \partial^{r_\ast} \Phi^\ast \right)
\propto
- 2 (\omega - m \Omega_H) |\Phi_{\mathrm{in}}|^2
\end{equation}
up to the same type of metric conversion factor that appeared in the Schwarzschild case. More explicitly, in the near-horizon limit one finds the same structural result
\begin{equation}
J^{r_\ast}
\sim
- \frac{2(\omega - m \Omega_H)}{\mathcal F_H} |\Phi_{\mathrm{in}}|^2
\end{equation}
where \(\mathcal F_H\) is the Kerr analogue of the Schwarzschild redshift factor, vanishing linearly at the horizon. The key point is not the exact prefactor but the shifted frequency and thus the sign and magnitude of the horizon flux are governed by
\begin{equation}
\omega - m \Omega_H
\end{equation}
rather than by \(\omega\) alone. Consequently the exterior quasilocal charge obeys
\begin{equation}
\frac{d}{dt} Q_{\mathrm{ext}}(t)
=
- \int_{\mathcal H_\epsilon} dS \, s_\mu J^\mu
\propto
(\omega - m \Omega_H)
\end{equation}
for a single near-horizon mode. This has three physically distinct regimes. First, if
\begin{equation}
\omega - m \Omega_H > 0
\end{equation}
then the horizon flux is inward in the same sense as in Schwarzschild and the exterior region loses quasilocal probability and the reduced exterior generator has an anti-Hermitian component corresponding to absorption by the black hole. Second, if
\begin{equation}
\omega - m \Omega_H = 0
\end{equation}
then the net horizon flux vanishes at leading order and in this threshold case, the horizon generator \(\chi^\mu\) sees no net flow of inner-product charge across the boundary and the exterior quasilocal charge is approximately conserved. Third, if
\begin{equation}
\omega - m \Omega_H < 0
\end{equation}
then the sign of the horizon flux reverses and in this case the exterior region gains quasilocal probability rather than losing it. This is precisely the regime associated with superradiant amplification \cite{sup1brito2015superradiance,sup2roy2022superradiance,sup3huang2019superradiant}. In the present language superradiance thus means that the horizon acts not as a sink of quasilocal probability but as an effective source for the restricted exterior description. The corresponding anti-Hermitian term in the reduced generator then has the opposite sign, describing effective gain rather than loss and this is the essential novelty introduced by rotation. In Schwarzschild, the horizon always behaves as an absorptive boundary for positive frequency ingoing modes while in Kerr the horizon can either absorb, remain neutral or amplify, depending on the sign of \(\omega - m \Omega_H\). Thus quasilocal probability in a rotating black hole background is not merely sensitive to the existence of a causal boundary, but also to the rotational state of the spacetime and to the azimuthal quantum number of the mode.
\\
\\
In Schwarzschild, quasilocal probability is the norm accessible outside a non-rotating horizon and its non-conservation reflects one way leakage into the black hole. In Kerr, quasilocal probability is the norm accessible outside a rotating horizon and its non-conservation reflects a competition between absorption and rotational amplification governed by \(\omega - m \Omega_H\). In the case of FLRW, quasilocal probability is the norm contained within a finite cosmological domain and its non-conservation reflects transport across the boundary weighted by the evolving scale factor. The same formal structure acquires three different geometric meanings which pertain to loss to a static horizon, exchange with a rotating horizon  and dilution or transport in an expanding universe. This comparison makes clear that quasilocal probability is not tied to any single special spacetime but is a general geometric refinement of probability conservation in curved backgrounds and what changes from one case to another is the way spacetime geometry determines the boundary through which the conserved current flows. Static horizons make the effect purely absorptive, rotating horizons make it frequency and angular momentum dependent and cosmological expansion makes it explicitly time dependent through the scale factor. In all cases, however, the full theory remains globally unitary when the complete domain is included while restricted observers naturally inherit an effective non-Hermitian description determined by geometric flux.
\\
\\
\section{Observational Imprints in Black Hole Ringdowns}
Quasilocal energy in general relativity provides a useful precedent for thinking about observational consequences of geometrically defined balance laws. For instance, in asymptotically flat spacetimes the ADM mass \cite{adm1Arnowitt:1959ah,adm2DeWitt:1967yk,adm3Arnowitt:1962hi}
\begin{equation}
M_{\text{ADM}} = \frac{1}{16\pi G} \int_{S_{\infty}} dS_i \left( \partial_j h_{ij} - \partial_i h_{jj} \right)
\end{equation}
influences orbital dynamics and gravitational wave emission, while in spherically symmetric cosmology the Misner-Sharp energy \cite{misner1964relativistic}
\begin{equation}
E_{\text{MS}} = \frac{R}{2G} \left( 1 - g^{\mu\nu} \partial_{\mu} R \partial_{\nu} R \right)
\end{equation}
controls collapse and expansion dynamics. Since quasilocal probability also arises from a geometric flux balance,
\begin{equation}
Q_R (t_2) - Q_R (t_1)
=
-\int_{B_R} d\Sigma_{\mu} J^{\mu}
\end{equation}
it is natural to ask whether measurable quantities sensitive to boundary flux could encode information about effective non-Hermiticity induced by spacetime structure.
\\
\\
The situation is, however, more indirect than for energy as global conservation of the inner-product current,
\begin{equation*}
\nabla_{\mu} J^{\mu} = 0
\end{equation*}
ensures that on a complete Cauchy surface without boundary flux the total charge remains constant and global Hermiticity is preserved. Observable effects therefore arise only when one considers restricted domains, for which
\begin{equation}
\frac{d}{dt} Q_R (t)
=
i \left( \Phi , ( H_R - H_R^{\dagger} ) \Phi \right)_R
\end{equation}
so that the anti-Hermitian part of the regional generator is controlled by boundary flux. Any observational imprint must therefore be encoded in quantities sensitive to effective damping, decoherence or absorption rates.
\\
\\
Black hole ringdowns \cite{qnm6silva2023black,qnm4baibhav2018black,qnm1nollert1996significance,qnm2berti2009quasinormal,qnm3konoplya2011quasinormal,qnm5mitman2023nonlinearities,qnm7isi2021analyzing,qnm8giesler2019black,abssanchez1978absorption} may provide the most immediate arena for such effects and the reason for that is structural. In the quasilocal framework developed here, the exterior region of a black hole spacetime does not admit an absolutely conserved inner-product charge by itself and for the exterior domain one has the balance law
\begin{equation}
\frac{dQ_{\rm ext}}{dt}=-\Phi_H
\label{eq:ringQbalance}
\end{equation}
where
\begin{equation}
Q_{\rm ext}=\int_{\Sigma_{\rm ext}} d\Sigma \, n_\mu J^\mu
\end{equation}
is the quasilocal probability outside the horizon and
\begin{equation}
\Phi_H=\int_{\mathcal H} dS \, J^\mu s_\mu
\end{equation}
is the flux of the inner-product current through the horizon. Since ringdown is precisely the regime in which the dynamics is governed by horizon and infinity boundary conditions, it is the cleanest observational setting in which this flux can manifest itself. The relevant observable is not only the oscillation frequency of the mode but especially its damping rate, because the latter is the direct imprint of norm leakage from the exterior sector into the inaccessible interior.
\\
\\
For linear perturbations of a stationary black hole spacetime one may write
\begin{equation}
\Psi(t,r,\theta,\phi)=\sum_{\ell m n} A_{\ell m n} e^{-i\omega_{\ell m n} t} \, {}_{-2}S_{\ell m}(\theta,\phi)\,\psi_{\ell m n}(r)
\label{eq:separation}
\end{equation}
where \({}_{-2}S_{\ell m}\) are the relevant spin weighted spheroidal harmonics and the radial function satisfies a master equation of Regge-Wheeler, Zerilli or Teukolsky type,
\begin{equation}
\left[\frac{d^2}{dr_*^2}+\omega^2-V_{\ell}(r)\right]\psi_{\ell m n}(r)=0
\label{eq:mastereq}
\end{equation}
with \(r_*\) being the tortoise coordinate as usual. In classical general relativity the quasi normal mode spectrum is defined by the boundary conditions
\begin{equation}
\psi(r_*\rightarrow -\infty)\sim e^{-i\omega r_*}
\qquad
\psi(r_*\rightarrow +\infty)\sim e^{+i\omega r_*}
\label{eq:grbc}
\end{equation}
corresponding respectively to purely ingoing waves at the horizon and purely outgoing waves at infinity and these conditions define a non self-adjoint spectral problem with complex eigenfrequencies
\begin{equation}
\omega_{\ell m n}^{(0)}=\omega_{R,\ell m n}^{(0)}+i\omega_{I,\ell m n}^{(0)}
\qquad
\omega_{I,\ell m n}^{(0)}<0
\label{eq:qnmcomplex}
\end{equation}
so that the mode amplitude decays as
\begin{equation}
\Psi_{\ell m n}(t)\propto e^{-i\omega_{R,\ell m n}^{(0)} t} e^{\omega_{I,\ell m n}^{(0)} t}
\label{eq:mode_decay}
\end{equation}
and the observed ringdown frequency and damping time are
\begin{equation}
f_{\ell m n}^{(0)}=\frac{\omega_{R,\ell m n}^{(0)}}{2\pi}
\qquad
\tau_{\ell m n}^{(0)}=\frac{1}{|\omega_{I,\ell m n}^{(0)}|}
\label{eq:freqtau0}
\end{equation}
Now what we should note is in the present framework, the horizon is not merely a classical absorber of gravitational wave energy but it is also a causal boundary through which inner-product charge flows. This means that the effective boundary condition at the horizon for the reduced exterior dynamics need not be exactly the classical one. A minimal parametrization is to then write
\begin{equation}
\left(\partial_{r_*}+i\omega\right)\psi\Big|_{H}
=
\varepsilon \, \mathcal K_H(\omega,\ell,m,n)\,\psi\Big|_{H}
\label{eq:modifiedBC}
\end{equation}
where \(\varepsilon\) is a small dimensionless parameter measuring the strength of the horizon induced obstruction of inner product conservation and \(\mathcal K_H\) is an effective response function encoding the detailed near horizon physics of the quasilocal flux. In the limit \(\varepsilon\to 0\) one recovers the standard purely ingoing condition and equation \eqref{eq:modifiedBC} is equivalent to leading order, to allowing a small deformation of the near-horizon mode content. The important point is that the deformation is localized at the horizon and therefore directly affects the quasi-normal spectrum. The resulting eigenfrequencies are determined by a spectral equation of the form
\begin{equation}
F_{\ell m n}(\omega,\varepsilon)=0
\label{eq:spectralcondition}
\end{equation}
with
\begin{equation}
F_{\ell m n}\left(\omega_{\ell m n}^{(0)},0\right)=0
\end{equation}
for the general relativistic solution and then we can write
\begin{equation}
\omega_{\ell m n}
=
\omega_{\ell m n}^{(0)}+\delta\omega_{\ell m n}
\label{eq:omegashiftdef}
\end{equation}
and expanding \eqref{eq:spectralcondition} to first order gives
\begin{equation}
0=
F_{\ell m n}\left(\omega_{\ell m n}^{(0)},0\right)
+
\left.\frac{\partial F_{\ell m n}}{\partial \omega}\right|_0 \delta\omega_{\ell m n}
+
\left.\frac{\partial F_{\ell m n}}{\partial \varepsilon}\right|_0 \varepsilon
+\mathcal O(\varepsilon^2)
\end{equation}
hence
\begin{equation}
\delta\omega_{\ell m n}
=
-
\varepsilon
\frac{\left.\partial_{\varepsilon}F_{\ell m n}\right|_0}
{\left.\partial_{\omega}F_{\ell m n}\right|_0}
+\mathcal O(\varepsilon^2)
\label{eq:deltaw_general}
\end{equation}
this shows explicitly that the quasinormal mode shifts are linear in the quasilocal probability parameter at leading order. Splitting \(\delta\omega_{\ell m n}\) into real and imaginary parts we can then write
\begin{equation}
\delta\omega_{\ell m n}
=
\delta\omega_{R,\ell m n}
+
i\,\delta\omega_{I,\ell m n}
\end{equation}
we obtain
\begin{equation}
\omega_{R,\ell m n}
=
\omega_{R,\ell m n}^{(0)}+\delta\omega_{R,\ell m n}
\qquad
\omega_{I,\ell m n}
=
\omega_{I,\ell m n}^{(0)}+\delta\omega_{I,\ell m n}
\label{eq:omegareim}
\end{equation}
and therefore
\begin{equation}
f_{\ell m n}
=
\frac{\omega_{R,\ell m n}}{2\pi}
=
f_{\ell m n}^{(0)}
\left(
1+\frac{\delta\omega_{R,\ell m n}}{\omega_{R,\ell m n}^{(0)}}
\right)
\label{eq:fshift}
\end{equation}
\begin{equation}
\tau_{\ell m n}
=
\frac{1}{|\omega_{I,\ell m n}|}
\simeq
\tau_{\ell m n}^{(0)}
\left(
1-\frac{\delta\omega_{I,\ell m n}}{\omega_{I,\ell m n}^{(0)}}
\right)
\label{eq:taushift}
\end{equation}
to first order in \(\varepsilon\). It is convenient to define dimensionless response coefficients
\begin{equation}
\alpha_{\ell m n}
\equiv
\frac{1}{\omega_{R,\ell m n}^{(0)}}
\Re\!\left[
-
\frac{\left.\partial_{\varepsilon}F_{\ell m n}\right|_0}
{\left.\partial_{\omega}F_{\ell m n}\right|_0}
\right]
\qquad
\beta_{\ell m n}
\equiv
-
\frac{1}{\omega_{I,\ell m n}^{(0)}}
\Im\!\left[
-
\frac{\left.\partial_{\varepsilon}F_{\ell m n}\right|_0}
{\left.\partial_{\omega}F_{\ell m n}\right|_0}
\right]
\label{eq:alphabeta}
\end{equation}
so that
\begin{equation}
\frac{\delta f_{\ell m n}}{f_{\ell m n}^{(0)}}=\alpha_{\ell m n}\varepsilon
\qquad
\frac{\delta \tau_{\ell m n}}{\tau_{\ell m n}^{(0)}}=\beta_{\ell m n}\varepsilon
\label{eq:freqtauepsilon}
\end{equation}
The crucial feature of \eqref{eq:freqtauepsilon} is that the same microscopic parameter \(\varepsilon\) induces correlated shifts across frequencies and damping times of different modes. Black hole spectroscopy is therefore not only sensitive to the magnitude of the effect but also to its specific multi mode pattern as well.
\\
\\
This can now be related directly to the quasilocal balance law and for that let \(a_{\ell m n}(t)\) denote the slowly varying mode amplitude for a given ringdown mode in the exterior region and let \(Q_{\ell m n}^{\rm ext}\) be the corresponding contribution to the exterior inner-product charge. Since the mode decays as \(a_{\ell m n}(t)\propto e^{-i\omega_{\ell m n} t}\), one has
\begin{equation}
Q_{\ell m n}^{\rm ext}(t)\propto |a_{\ell m n}(t)|^2
\propto e^{2\omega_{I,\ell m n} t}
\end{equation}
and therefore
\begin{equation}
\frac{dQ_{\ell m n}^{\rm ext}}{dt}=2\omega_{I,\ell m n}Q_{\ell m n}^{\rm ext}
\label{eq:dQmode}
\end{equation}
Comparing this with the quasilocal balance law \eqref{eq:ringQbalance} we obtain the identification
\begin{equation}
2\omega_{I,\ell m n}Q_{\ell m n}^{\rm ext}=-\Phi_{H,\ell m n}
\end{equation}
or equivalently
\begin{equation}
\omega_{I,\ell m n}
=
-\frac{\Phi_{H,\ell m n}}{2Q_{\ell m n}^{\rm ext}}
\label{eq:omegai_flux}
\end{equation}
This equation is central for us and we should emphasize why. This shows that the imaginary part of the quasinormal mode frequency is nothing but the quasilocal probability leakage rate per unit exterior charge. Hence, among all possible strong gravity observables, the ringdown damping time is the most direct probe of quasilocal probability.
\\
\\
The corresponding gravitational wave strain in the ringdown regime may be written as
\begin{equation}
h(t)=\sum_{\ell m n} \mathcal A_{\ell m n}
e^{-t/\tau_{\ell m n}}
\cos\!\left(2\pi f_{\ell m n} t+\phi_{\ell m n}\right)\Theta(t-t_0)
\label{eq:ringdownwaveform}
\end{equation}
where \(t_0\) is the ringdown start time and in the present framework one may parameterize the leading quasilocal corrections as
\begin{equation}
f_{\ell m n}=f_{\ell m n}^{\rm Kerr}\left(1+\alpha_{\ell m n}\varepsilon\right)
\qquad
\tau_{\ell m n}=\tau_{\ell m n}^{\rm Kerr}\left(1+\beta_{\ell m n}\varepsilon\right)
\qquad
\mathcal A_{\ell m n}=\mathcal A_{\ell m n}^{\rm Kerr}\left(1+\gamma_{\ell m n}\varepsilon\right)
\label{eq:parametrization}
\end{equation}
where \(\gamma_{\ell m n}\) allows for a small correction in mode excitation. Substituting \eqref{eq:parametrization} into \eqref{eq:ringdownwaveform} and expanding to first order in \(\varepsilon\) gives
\begin{equation}
h(t)=h_{\rm Kerr}(t)+\varepsilon\,\delta h(t)+\mathcal O(\varepsilon^2)
\label{eq:hlin}
\end{equation}
with
\begin{equation}
\delta h(t)=
\sum_{\ell m n}
\left[
\gamma_{\ell m n}\mathcal A_{\ell m n}^{\rm Kerr}\frac{\partial h}{\partial \mathcal A_{\ell m n}}
+
\alpha_{\ell m n} f_{\ell m n}^{\rm Kerr}\frac{\partial h}{\partial f_{\ell m n}}
+
\beta_{\ell m n} \tau_{\ell m n}^{\rm Kerr}\frac{\partial h}{\partial \tau_{\ell m n}}
\right]_{\rm Kerr}
\label{eq:deltah}
\end{equation}
For a single damped sinusoid we then have
\begin{equation}
h_n(t)=\mathcal A_n e^{-t/\tau_n}\cos(2\pi f_n t+\phi_n)\Theta(t-t_0)
\label{eq:singlemode}
\end{equation}
the needed derivatives are
\begin{equation}
\frac{\partial h_n}{\partial \mathcal A_n}
=
e^{-t/\tau_n}\cos(2\pi f_n t+\phi_n)\Theta(t-t_0)
\end{equation}
\begin{equation}
\frac{\partial h_n}{\partial f_n}
=
-2\pi t\,\mathcal A_n e^{-t/\tau_n}\sin(2\pi f_n t+\phi_n)\Theta(t-t_0)
\end{equation}
\begin{equation}
\frac{\partial h_n}{\partial \tau_n}
=
\mathcal A_n e^{-t/\tau_n}\frac{t}{\tau_n^2}\cos(2\pi f_n t+\phi_n)\Theta(t-t_0)
\end{equation}
and therefore
\begin{equation}
\frac{\partial h_n}{\partial \varepsilon}
=
\gamma_n \mathcal A_n^{\rm Kerr}\frac{\partial h_n}{\partial \mathcal A_n}
+
\alpha_n f_n^{\rm Kerr}\frac{\partial h_n}{\partial f_n}
+
\beta_n \tau_n^{\rm Kerr}\frac{\partial h_n}{\partial \tau_n}
\label{eq:dhdvareps}
\end{equation}
This form is particularly useful for a Fisher matrix analysis as for that , consider that we have detector noise power spectral density \(S_n(f)\), after which the standard noise weighted inner product is
\begin{equation}
(a|b)=4\,\Re\int_0^\infty df \,
\frac{\tilde a(f)\tilde b^*(f)}{S_n(f)}
\label{eq:noiseproduct}
\end{equation}
If \(\theta^a\) denotes the waveform parameters including \(\varepsilon\), then the Fisher matrix is
\begin{equation}
\Gamma_{ab}=
\left(
\frac{\partial h}{\partial \theta^a}
\Bigg|
\frac{\partial h}{\partial \theta^b}
\right)
\label{eq:fisher}
\end{equation}
and the expected \(1\sigma\) uncertainty on \(\varepsilon\) is
\begin{equation}
\sigma_\varepsilon\simeq \sqrt{\left(\Gamma^{-1}\right)_{\varepsilon\varepsilon}}
\label{eq:sigmaeps}
\end{equation}
In the idealized case where correlations with the other parameters are mild, this reduces to
\begin{equation}
\sigma_\varepsilon \simeq \Gamma_{\varepsilon\varepsilon}^{-1/2}
\qquad
\Gamma_{\varepsilon\varepsilon}
=
\left(
\frac{\partial h}{\partial \varepsilon}
\Bigg|
\frac{\partial h}{\partial \varepsilon}
\right)
\label{eq:simplefisher}
\end{equation}
Using \eqref{eq:dhdvareps} one then obtains
\begin{equation}
\Gamma_{\varepsilon\varepsilon}
\sim
\sum_n \rho_n^2
\left(
c_n^{(A)}\gamma_n^2
+
c_n^{(f)}\alpha_n^2
+
c_n^{(\tau)}\beta_n^2
+
{\rm cross\ terms}
\right)
\label{eq:fisherscaling}
\end{equation}
where \(\rho_n\) is the ringdown signal to noise ratio in mode \(n\) and the coefficients \(c_n^{(A)},c_n^{(f)},c_n^{(\tau)}\) are dimensionless numbers of order unity that depend on the damping rate, start time and detector sensitivity curve. Equation \eqref{eq:fisherscaling} immediately implies to us the characteristic scaling
\begin{equation}
\sigma_\varepsilon \propto \rho_{\rm rd}^{-1}
\label{eq:snrscaling}
\end{equation}
This ends up showing that ringdown is not only conceptually clean but also observationally efficient as every improvement in post merger signal-to-noise translates directly into a tighter constraint on quasilocal probability.
\\
\\
We now estimate what current data imply and for that we use gravitational wave events. The loud event GW250114 \cite{ob1LIGOScientific:2025wao} has provided the strongest single event ringdown constraints to date and in a full inspiral-merger-ringdown spectroscopy analysis, the dominant \((\ell,m,n)=(2,2,0)\) mode was found to be consistent with Kerr with fractional deviations \cite{ob2Ghosh:2021mrv}
\begin{equation}
\delta \hat f_{220}=0.02^{+0.02}_{-0.02}
\qquad
\delta \hat \tau_{220}=-0.01^{+0.10}_{-0.09}
\label{eq:currentbounds220}
\end{equation}
at \(90\%\) credibility while the combined GWTC-4.0 result gave the weaker bounds
\begin{equation}
\delta \hat f_{220}=0.00^{+0.06}_{-0.06}
\qquad
\delta \hat \tau_{220}=0.16^{+0.18}_{-0.16}
\label{eq:gwtc4bounds220}
\end{equation}
For the first overtone, one obtains constraints at the level of tens of percent for example
\begin{equation}
\delta \hat f_{221}=0.09^{+0.29}_{-0.30}
\label{eq:221bounds}
\end{equation}
for one representative fit choice, again at \(90\%\) credibility. The \((4,4,0)\) mode is also beginning to be constrained, with
\begin{equation}
\delta \hat f_{440}=-0.06^{+0.25}_{-0.35}
\qquad
\delta \hat \tau_{440}=0.20^{+0.53}_{-0.69}
\label{eq:440bounds}
\end{equation}
These numbers show two things at once wherein first, the dominant mode is already measured at the percent level in frequency and at the ten percent level in damping time and second, subdominant modes are not yet measured comparably well, meaning that multimode correlated tests of the type predicted here are only beginning to become observationally viable.
\\
\\
To translate the dominant mode bounds into a constraint on quasilocal probability, we identify
\begin{equation}
\left|\frac{\delta f_{220}}{f_{220}^{(0)}}\right| \simeq |\alpha_{220}\varepsilon|
\qquad
\left|\frac{\delta \tau_{220}}{\tau_{220}^{(0)}}\right| \simeq |\beta_{220}\varepsilon|
\label{eq:epsboundbasic}
\end{equation}
Using \eqref{eq:currentbounds220}, the \(90\%\) credibility scales are approximately
\begin{equation}
\left|\frac{\delta f_{220}}{f_{220}^{(0)}}\right| \lesssim 2\times 10^{-2}
\qquad
\left|\frac{\delta \tau_{220}}{\tau_{220}^{(0)}}\right| \lesssim 10^{-1}
\label{eq:220sizes}
\end{equation}
Hence we have
\begin{equation}
|\varepsilon|
\lesssim
\min\left(
\frac{2\times 10^{-2}}{|\alpha_{220}|},
\frac{10^{-1}}{|\beta_{220}|}
\right)
\label{eq:epsboundmin}
\end{equation}
If the response coefficients are of order unity, \(|\alpha_{220}|\sim |\beta_{220}|\sim 1\), then the current GW250114 data imply
\begin{equation}
|\varepsilon| \lesssim 10^{-2}\mbox{-}10^{-1}
\label{eq:epsorderofmag}
\end{equation}
with the stronger end of this range driven by the frequency measurement of the dominant mode and the more conservative end by the damping time measurement. Since the quasilocal framework naturally predicts that the imaginary part of the spectrum is the most direct manifestation of horizon induced probability leakage, the conservative and conceptually most robust statement is
\begin{equation}
|\varepsilon| \lesssim 10^{-1}
\label{eq:epsconservative}
\end{equation}
for the presently accessible astrophysical ringdown sector, while the best current single event data already begin to probe
\begin{equation}
|\varepsilon| \sim {\rm few}\times 10^{-2}
\label{eq:epsoptimistic}
\end{equation}
under the natural assumption of \(\mathcal O(1)\) response coefficients.
\\
\\
This is an important conclusion as this shows that current ringdown observations do not rule out quasilocal probability at all. Rather, they imply that any horizon-induced obstruction of inner-product conservation for the exterior observer must be perturbatively small. In other words, the data indicate that canonical Hermiticity remains an excellent approximation for the exterior algebra of presently observed black holes, but they do not require it to be exact. The framework therefore survives current tests comfortably. One can also see this from the measured values of the dominant mode itself as the GW250114 spectroscopy analysis finds
\begin{equation}
f_{220}=251.7^{+5.1}_{-5.0}\,{\rm Hz}
\qquad
\tau_{220}=4.09^{+0.42}_{-0.38}\,{\rm ms}
\label{eq:actual220values}
\end{equation}
which correspond to relative uncertainties of about
\begin{equation}
\frac{\Delta f_{220}}{f_{220}}\sim 2\times 10^{-2}
\qquad
\frac{\Delta \tau_{220}}{\tau_{220}}\sim 10^{-1}
\label{eq:relative220unc}
\end{equation}
exactly matching the order-of-magnitude estimate above. The event’s network signal-to-noise ratio was about \(76\), with post merger signal-to-noise ratio around \(40\), and this immediately explains to us why the dominant mode is already tightly constrained while subdominant modes remain much less so. The Fisher scaling \eqref{eq:snrscaling} then tells us that future louder ringdowns and cleaner multimode measurements should improve the bound on \(\varepsilon\) roughly inversely with ringdown signal-to-noise given all else is equal.
\\
\\
Finally, ringdown is not merely one possible probe among many but it is the best near term probe of quasilocal probability because the fundamental effect enters exactly at the level where the observable is defined. In the present framework the horizon flux modifies the exterior boundary condition, the boundary condition shifts the quasi-normal spectrum, and the quasinormal spectrum is what ringdown spectroscopy directly measures. The logical chain is therefore exceptionally short and we can summarize that as follows:
\begin{equation}
\Phi_H \neq 0
\quad \Longrightarrow \quad
\mbox{modified horizon boundary condition}
\quad \Longrightarrow \quad
\delta \omega_{\ell m n}\neq 0
\quad \Longrightarrow \quad
\delta f_{\ell m n},\,\delta \tau_{\ell m n}\neq 0
\label{eq:logicalchain}
\end{equation}
This directness is what makes black hole ringdown the most natural observational arena for quasilocal probability effects and the present data already constrain the associated parameter \(\varepsilon\) to be small, but by no means zero and therefore leave the framework very much alive. As multimode black hole spectroscopy matures, it will either tighten these bounds significantly or reveal the correlated departures from Kerr ringdown that a genuinely quasilocal notion of probability predicts.
\\
\\
A natural question especially from the perspective of gravitational wave data analysis could be the following ; what does the quasilocal probability framework predict directly for the observed ringdown strain $h(t)$ relative to the standard Kerr expectation? To answer this, we begin by noting that in the ringdown regime, the strain can be written as a sum of damped quasinormal modes,
\begin{equation}
h(t)=\sum_{\ell mn} A_{\ell mn} e^{-t/\tau_{\ell mn}} \cos\!\left(2\pi f_{\ell mn} t+\phi_{\ell mn}\right)\Theta(t-t_{0})
\end{equation}
where $t_{0}$ denotes the ringdown start time and in the present framework, the leading quasilocal corrections are already encoded through the mode-parameter shifts
\begin{equation}
f_{\ell mn}=f^{\rm Kerr}_{\ell mn}(1+\alpha_{\ell mn}\epsilon), \qquad
\tau_{\ell mn}=\tau^{\rm Kerr}_{\ell mn}(1+\beta_{\ell mn}\epsilon), \qquad
A_{\ell mn}=A^{\rm Kerr}_{\ell mn}(1+\gamma_{\ell mn}\epsilon)
\end{equation}
where $\epsilon$ is the small dimensionless quasilocal probability parameter and $\alpha_{\ell mn}$, $\beta_{\ell mn}$ and $\gamma_{\ell mn}$ are response coefficients again and this immediately gives
\begin{equation}
h(t)=h_{\rm Kerr}(t)+\epsilon\,\delta h(t)+\mathcal{O}(\epsilon^{2})
\end{equation}
with
\begin{equation}
\delta h(t)=\sum_{\ell mn}\left[
\gamma_{\ell mn}A^{\rm Kerr}_{\ell mn}\frac{\partial h}{\partial A_{\ell mn}}
+\alpha_{\ell mn}f^{\rm Kerr}_{\ell mn}\frac{\partial h}{\partial f_{\ell mn}}
+\beta_{\ell mn}\tau^{\rm Kerr}_{\ell mn}\frac{\partial h}{\partial \tau_{\ell mn}}
\right]_{\rm Kerr}
\end{equation}
For a single mode, which is sufficient for an illustrative comparison with the standard Kerr prediction, one has
\begin{equation}
h_{n}^{\rm Kerr}(t)=A_{n}^{\rm Kerr} e^{-t/\tau_{n}^{\rm Kerr}}
\cos\!\left(2\pi f_{n}^{\rm Kerr} t+\phi_{n}\right)\Theta(t-t_{0})
\end{equation}
and therefore the quasilocal probability corrected waveform takes the explicit form
\begin{equation}
h_{n}^{\rm QL}(t)=A_{n}^{\rm Kerr}(1+\gamma_{n}\epsilon)
\exp\!\left[-\frac{t}{\tau_{n}^{\rm Kerr}(1+\beta_{n}\epsilon)}\right]
\cos\!\left(2\pi f_{n}^{\rm Kerr}(1+\alpha_{n}\epsilon)t+\phi_{n}\right)\Theta(t-t_{0})
\end{equation}
To first order in $\epsilon$, this becomes
\begin{equation}
h_{n}^{\rm QL}(t)\approx h_{n}^{\rm Kerr}(t)+\epsilon\,A_{n}^{\rm Kerr}e^{-t/\tau_{n}^{\rm Kerr}}
\left[
\gamma_{n}\cos\Psi_{n}
-2\pi \alpha_{n} f_{n}^{\rm Kerr} t \sin\Psi_{n}
+\beta_{n}\frac{t}{\tau_{n}^{\rm Kerr}}\cos\Psi_{n}
\right]\Theta(t-t_{0})
\end{equation}
where
\begin{equation}
\Psi_{n}=2\pi f_{n}^{\rm Kerr} t+\phi_{n}
\end{equation}
This expression makes the observational content more transparent, as the $\alpha_{n}$ term produces a phase drift relative to Kerr, the $\beta_{n}$ term modifies the damping envelope and the $\gamma_{n}$ term changes the mode excitation amplitude. Since the quasilocal probability framework ties the anti-Hermitian contribution most directly to horizon flux, the most characteristic effect is a shift in the damping time
\begin{equation}
\tau_{n}^{\rm QL}=\tau_{n}^{\rm Kerr}(1+\beta_{n}\epsilon)
\end{equation}
so that
\begin{equation}
\frac{\delta \tau_{n}}{\tau_{n}^{\rm Kerr}}=\beta_{n}\epsilon
\end{equation}
and equivalently
\begin{equation}
\frac{\delta f_{n}}{f_{n}^{\rm Kerr}}=\alpha_{n}\epsilon
\end{equation}
Thus a detection of quasilocal probability in ringdown would not appear as a qualitatively new waveform morphology, but rather as a systematic and mode-dependent departure from the Kerr values of the ringdown frequencies, damping times and possibly amplitudes. For the dominant mode, the waveform level effect may be visualized simply by comparing
\begin{equation}
h_{220}^{\rm Kerr}(t)=A_{220}e^{-t/\tau^{\rm Kerr}_{220}}
\cos\!\left(2\pi f^{\rm Kerr}_{220}t+\phi_{220}\right)
\end{equation}
with
\begin{equation}
h_{220}^{\rm QL}(t)=A_{220}(1+\gamma_{220}\epsilon)
\exp\!\left[-\frac{t}{\tau^{\rm Kerr}_{220}(1+\beta_{220}\epsilon)}\right]
\cos\!\left(2\pi f^{\rm Kerr}_{220}(1+\alpha_{220}\epsilon)t+\phi_{220}\right)
\end{equation}
for representative values of $\epsilon$. A positive $\beta_{220}\epsilon$ lengthens the damping time and leads to a more slowly decaying ringdown than Kerr, while a negative $\beta_{220}\epsilon$ gives a faster decay. In multimode spectroscopy, the more distinctive signature is not merely a shift in one dominant mode, but a correlated pattern across several $(\ell,m,n)$ channels,
\begin{equation}
\left\{\frac{\delta f_{\ell mn}}{f_{\ell mn}^{\rm Kerr}},\,
\frac{\delta \tau_{\ell mn}}{\tau_{\ell mn}^{\rm Kerr}},\,
\frac{\delta A_{\ell mn}}{A_{\ell mn}^{\rm Kerr}}\right\}
=
\left\{\alpha_{\ell mn}\epsilon,\,
\beta_{\ell mn}\epsilon,\,
\gamma_{\ell mn}\epsilon\right\}
\end{equation}
all controlled by the same underlying parameter $\epsilon$. This is the sense in which the present framework predicts a structured deviation from Kerr rather than an arbitrary phenomenological waveform deformation.
\begin{figure}[!h]
    \centering
    \includegraphics[width=1\linewidth]{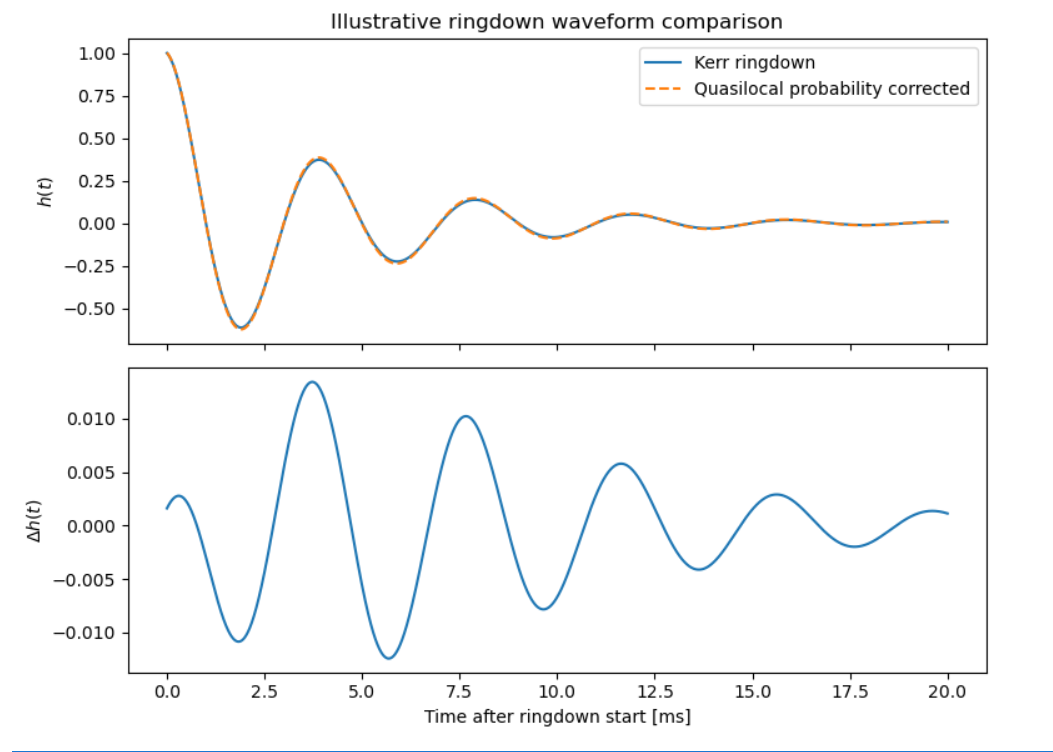}
    \caption{Illustrative comparison of the standard Kerr ringdown waveform and the quasilocal probability corrected waveform for the dominant mode. The quasilocal model produces small correlated shifts in the frequency, damping time and amplitude, parameterized by $\alpha_{220}\epsilon$, $\beta_{220}\epsilon$ and $\gamma_{220}\epsilon$. In this example the dominant visible effect is a small change in the damping envelope together with an accumulating phase drift relative to Kerr.}
\label{fig:placeholder}
\end{figure}
Illustrative comparison of the standard Kerr ringdown waveform and the quasilocal probability corrected waveform for the dominant mode has been done in figure \ref{fig:placeholder}. The Kerr waveform (solid blue) is given by 
$h_{\rm Kerr}(t)=A e^{-t/\tau} \cos(2\pi f t+\phi)$, while the quasilocal waveform (dashed orange) incorporates the leading-order corrections derived in this work,
$f = f^{\rm Kerr}(1+\alpha\epsilon)$, $\tau = \tau^{\rm Kerr}(1+\beta\epsilon)$ and $A = A^{\rm Kerr}(1+\gamma\epsilon)$, with $\epsilon$ the quasilocal probability parameter and for illustration, we adopt representative values $f^{\rm Kerr}=250\,\mathrm{Hz}$, $\tau^{\rm Kerr}=4\,\mathrm{ms}$ and small response coefficients $(\alpha,\beta,\gamma)=(0.03,0.40,0.02)$ with $\epsilon=0.08$, consistent with the perturbative regime implied by the flux-balance relations of Section 6. The top panel shows that the modified waveform closely tracks the Kerr prediction, reflecting the smallness of the effect while the bottom panel displays the residual $\Delta h(t)=h_{\rm QL}(t)-h_{\rm Kerr}(t)$. The oscillatory structure of the residual encodes a cumulative phase shift arising from $\alpha\epsilon$, while the slowly varying envelope reflects the change in damping time controlled by $\beta\epsilon$, which is directly tied to the quasilocal probability flux across the horizon. This goes to show that the primary observational signature of the framework is a correlated, mode dependent deviation in ringdown frequency and damping time rather than a qualitatively new waveform morphology. We also note that the magnitude of the residual is at the percent level for representative parameters, consistent with current ringdown sensitivities. In addition to this, we would like to share that in an upcoming follow-uo work on quasilocal probability, we would be discussing three smoking gun observations for quasilocal probability with black hole spectroscopy.  

\section{Conclusions}
In this work we have extended the idea of Hermiticity as a symmetry of inner-product conservation to curved spacetimes and shown that probability conservation acquires a fundamentally quasilocal character in the presence of gravitational structure. Starting from the interpretation of Hermiticity as the condition ensuring invariance of the quantum inner product, we demonstrated that in field theory this invariance is encoded in the divergence-free inner-product current. When spacetime contains intrinsic boundaries such as horizons or when observers are restricted to finite domains, the associated charge obeys a quasilocal flux balance law. We explicitly derived the form of this quasilocal probability in Schwarzschild and FLRW spacetimes, showing how metric components and geometric measure factors enter directly into its definition and how boundary flux induces effective non-Hermiticity in regional Hamiltonians. In both cases we verified that global Hermiticity is restored when the full Cauchy surface is included or when the flat spacetime limit is taken, thereby preserving the consistency of the underlying quantum theory while revealing a geometrically controlled mechanism for effective non-unitarity in restricted settings. We also showed how this framework can be observationally tested with black hole ringdown and spectroscopy.
\\
\\
But what does it all mean in the grand scheme of things? If probability, like energy, becomes quasilocal in curved spacetime, then the standard formulation of quantum mechanics as globally unitary evolution generated by a single Hermitian Hamiltonian must be viewed as a special limit tied to specific geometric conditions. In quantum field theory in curved spacetime, effective non-Hermiticity may naturally arise whenever horizons, expanding backgrounds, or restricted algebras prevent global implementation of inner-product symmetry. This suggests that open-system methods, flux balance laws, and geometry dependent inner-product structures should be incorporated more fundamentally into our treatment of quantum fields in gravitational backgrounds. At the interface of quantum gravity, where spacetime itself becomes dynamical and causal structure may fluctuate, quasilocal probability may provide a guiding principle for understanding how unitarity is realized or reinterpreted in regimes beyond semiclassical approximation. Developing a consistent framework in which probability balance, Hermiticity symmetry, and gravitational dynamics are treated on equal footing could lead to conceptual refinements in both quantum gravity and cosmology. We will explore these regimes and their structural implications for quantum field theory and quantum gravity in forthcoming works.
\\
\\
Quasilocal probability can be understood most simply by thinking about what an observer can actually access in spacetime. In ordinary quantum mechanics in flat spacetime, probability is treated as a global quantity, which means that if one integrates the probability density of a quantum state over all space, the result remains constant in time. For an observer who has access to the entire spatial slice of the universe, probability is therefore a strictly conserved quantity and the generator of time evolution must be Hermitian. In curved spacetime, however, observers rarely have access to the entire spacetime manifold and horizons, causal boundaries and finite domains limit the region from which signals can reach them. In such circumstances the observer can only measure the probability contained within the region that is causally accessible and thus quasilocal probability is precisely the statement that the normalization of the quantum state inside that finite region need not remain constant by itself, because probability can flow across the boundary of the region.
\\
\\
A useful Gedanken experiment illustrates this point more clearly and for this, imagine an observer hovering outside a black hole while monitoring a quantum field in the exterior region. From the perspective of this observer, waves of the field can propagate inward and eventually cross the event horizon. Once this happens the information carried by those modes becomes permanently inaccessible to the exterior observer. If the observer computes the probability contained in the region outside the horizon, they will find that it gradually decreases in time because part of the probability current flows into the black hole. From the global point of view nothing pathological has happened since the full spacetime including the interior still conserves probability, but for the restricted observer the quantum evolution appears effectively non-Hermitian, because the accessible Hilbert space is open to flux through the horizon. Quasilocal probability therefore provides a precise way to quantify how gravitational boundaries alter the operational meaning of probability.
\\
\\
The situation contrasts sharply with the familiar case of flat spacetime as consider an observer in Minkowski space who analyzes a quantum wave packet evolving in an otherwise empty universe. In the absence of horizons or intrinsic causal boundaries, the observer can in principle extend their region of observation arbitrarily far and if the wave packet spreads outward, the probability that leaves one region simply reappears in another region that the observer can still include in their description. By enlarging the domain of integration to cover the entire spatial slice, the observer always recovers exact conservation of the total probability. In this case, the boundary flux vanishes at infinity and the standard formulation of quantum mechanics with a globally conserved inner product emerges naturally.
\\
\\
From this perspective, quasilocal probability represents a conceptual shift in how quantum theory and spacetime geometry interact. Rather than viewing probability conservation as an absolute global statement, it becomes a geometric balance law that depends on the causal structure accessible to a given observer. Horizons, rotation and cosmological expansion determine how probability currents cross the boundaries of observable regions, thereby shaping the effective quantum dynamics seen by those observers. In curved spacetimes the quantum description therefore becomes inherently relational as what appears as perfectly unitary evolution globally may manifest locally as an open system whose normalization evolves through geometric flux across spacetime boundaries. One can only begin to imagine the amount of aftereffects of this on quantum field theory in curved spacetime then.
\\
\\
To understand quasilocal probability in an even more easier fundamental sense, think of a box filled with quantum particles, where the probability represents how much of the system is inside the box. Now, if the box is perfectly sealed, the total probability inside stays constant but if the box has a small hole in it, particles can leak out. So, the probability inside the box decreases even though nothing is destroyed globally. Now, one can think of spacetime regions like outside a black hole such as a "box", where horizons act like one way boundaries. An observer restricted to that region only sees the probability inside, which can change due to this leakage. So, what looks like loss of probability is really just loss of access to information beyond the boundary and this is exactly what is quasilocal probability and it naturally leads to effective non-Hermitian behavior. In the most intuitive way, quasilocal probability is just probability seen by an observer who cannot see the whole universe and thus gravity, fundmentally influences the way one sees probability itself. It doesn't at all mean that "quantum mechanics breaks", but what this really shows is that the observer's description becomes incomplete in the presence of horizons and spacetime curvature, and hence the mathematics adapts accordingly.
\section*{Acknowledgements}
Work of the author is supported in part by the Vanderbilt Discovery Doctoral Fellowship. The author would like to thank Robert Scherrer, Alfredo Gurrola, Abraham Loeb and Sunny Vagnozzi for various helpful discussions on the topics discussed in this work. The author also thanks Florencia Aravena (Floppy) for many insightful conversations and support. I would also like to thank the anonymous reviewer for their insightful comments on the paper.

\bibliography{references}
\bibliographystyle{unsrt}

\end{document}